\documentclass{article}
\usepackage{spconf}
\usepackage[hidelinks]{hyperref}
\usepackage[sort]{cite}
\usepackage{pbalance}
\usepackage{amsmath}
\usepackage{graphicx}
\usepackage{bm}
\usepackage{bbm}
\usepackage{amssymb}
\usepackage{xspace}
\usepackage{cite}
\usepackage{siunitx}
\usepackage{url}
\usepackage{booktabs}

\newcommand{\tsixty}{$\mathrm{T}_{60}$\xspace}
\newcommand{\tthirty}{$\mathrm{T}_{30}$\xspace}
\newcommand{\cfifty}{$\mathrm{C_{50}}$\xspace}
\DeclareMathOperator*{\argmin}{arg\,min}

\title{Matching Reverberant Speech Through Learned Acoustic Embeddings and Feedback Delay Networks}
\name{Philipp G\"{o}tz$^{1}$, Gloria Dal Santo$^{2}$, Sebastian J. Schlecht$^{3}$, Vesa V{\"a}lim{\"a}ki$^{2}$, Emanu\"{e}l A. P. Habets$^{1}$\thanks{$^\dag$A joint institution of the Friedrich-Alexander-Universit\"{a}t Erlangen-N\"{u}rnberg (FAU) and Fraunhofer IIS.}}
\address{$^1$ International Audio Laboratories Erlangen$^\dag$, Germany\\ 
$^{2}$ Acoustics Lab, Dpt. of Information and Communications Engineering, Aalto University, Finland\\
$^{3}$ Friedrich-Alexander-Universit\"{a}t Erlangen-N\"{u}rnberg (FAU), Germany}

\begin{document}
\ninept

\maketitle

\begin{abstract}
Reverberation conveys critical acoustic cues about the environment, supporting spatial awareness and immersion. For auditory augmented reality (AAR) systems, generating perceptually plausible reverberation in real time remains a key challenge, especially when explicit acoustic measurements are unavailable. We address this by formulating blind estimation of artificial reverberation parameters as a reverberant signal matching task, leveraging a learned room-acoustic prior. Furthermore, we propose a feedback delay network (FDN) structure that reproduces both frequency-dependent decay times and the direct-to-reverberation ratio of a target space. Experimental evaluation against a leading automatic FDN tuning method demonstrates improvements in estimated room-acoustic parameters and perceptual plausibility of artificial reverberant speech. These results highlight the potential of our approach for efficient, perceptually consistent reverberation rendering in AAR applications.
\end{abstract}

\section{Introduction}
\label{sec:intro}
Reverberation provides the auditory system with rich information about the size, geometry, and material properties of an environment, allowing listeners to form a sense of space and situational awareness \cite{blauert1997spatial,bregman1994auditory}. Accordingly, auditory augmented reality (AAR) systems must ensure that virtual sound sources integrate seamlessly into real acoustic settings to preserve immersion, realism, and telepresence \cite{agrawal2019defining,neidhardt2022perceptual,potter2022relative,immohr2023proof}. Plausible acoustic rendering is a core challenge in AAR, requiring a method to generate artificial reverberation in real-time while adhering to computational complexity constraints, as well as knowledge of the acoustic surroundings of the user \cite{meyer2024testing}.

Such knowledge about the acoustic environment may be represented by a room impulse response (RIR), from which room-acoustic parameters such as reverberation time \tsixty and clarity index \cfifty can be derived \cite{kuttruff2016room}, or through geometric information like the shape of the environment and the positions of sources and receivers. In many practical situations, acquiring such information via dedicated measurements or user input is infeasible, requiring instead a non-intrusive inference from observed reverberant signals \cite{eaton2016estimation}. Blind estimation of RIRs from reverberant signals, in the single- and multi-channel case, has long been an active area of research \cite{lin2006bayesian,crammer2006room}. Recent advances have been driven by leveraging the powerful discriminative and generative modeling capabilities of deep neural networks (DNNs) \cite{steinmetz2021filtered, lee2023yet,ratnarajah2023towards,moliner2024buddy}. However, many of these methods are computationally demanding and not specifically tailored towards applications that involve real-time processing of audio, such as AAR. Moreover, rendering virtual sound sources directly via convolution with estimated RIRs can be challenging as it is computationally expensive, requires storing or recomputing long filters for dynamic scenes, and lacks flexibility when the environment, source position, or listener position changes. Parametric artificial reverberation methods, by contrast, offer a lightweight alternative that enables real-time rendering while maintaining perceptual plausibility. Among these, feedback delay networks (FDNs) have emerged as a particularly effective and versatile class of structures for interactive applications \cite{schlecht2016feedback}.

FDNs were introduced as a generalization of the parallel comb-filter structure in which the delays are interconnected via a feedback matrix \cite{gerzon1971synthetic, jot1991digital}. The continued popularity of FDNs is due to their computational efficiency and flexibility, allowing real-time processing and independent control over parameters such as energy decay, diffusion, and overall equalization \cite{valimaki2012fifty}.
By reformulating the FDN in a differentiable form, DNNs can be used to estimate the parameter values needed to synthesize a target reverberation, thereby enabling blind estimation of RIRs. This idea has been explored by Sungho Lee et al. in \cite{lee2022differentiable}, where an artificial reverberator parameter estimation network (ARP-net) is used to determine a subset of FDN parameters from reverberant speech in an end-to-end manner. The ARP-net employs an encoder to convert audio spectrograms into a latent vector, followed by ARP-groupwise layers for FDN parameter projection. The model is particularly large, with the ARP-net comprising approximately 7.3M parameters. While the method has shown promising results, the chosen FDN design and optimization scheme, specifically the use of a fixed mixing matrix and a shared attenuation filter, limit further improvements and reduce perceptual plausibility \cite{dal2024rir2fdn}.

The main contributions of the present study are twofold. First, we formulate the blind estimation of artificial reverberation parameters as a reverberant signal matching task, utilizing a pre-learned room-acoustic prior. 
Second, we propose a differentiable FDN structure that, unlike prior work, offers greater flexibility in frequency-dependent decay and energy control, as well as improved temporal density. 
We evaluate the proposed approach against a leading automatic FDN tuning method in \cite{lee2022differentiable}, comparing performance in terms of room-acoustic parameters, specifically $T_{60}$ and $C_{50}$, as well as the perceptual plausibility of the resulting artificial reverberant speech reflected by the Fréchet Audio Distance (FAD) \cite{kilgour2019frechet}.

The remainder of this paper is structured as follows. Section~\ref{sec:method} introduces the method, extraction of a room-acoustic embedding from speech, as well as the differentiable FDN structure. Section~\ref{sec:eval} presents the evaluation setup and the baseline used for performance comparison. Section~\ref{sec:results} contains the presentation and discussion of experimental results, and Section~\ref{sec:conclusion} concludes this work.
\section{Proposed Method}

\label{sec:method}
We consider the reverberant signal $y[t]$ as the convolution of an anechoic source signal $x[t]$ with a RIR $h[t]$ of length $L$, where the result is corrupted by uncorrelated, additive background and sensor noise $v[t]$:
\begin{equation}
    y[t] = (x * h)[t] + v[t] =\sum_{\tau=0}^{L-1} x[t-\tau]h[\tau] + v[t],
\end{equation}
where $t$ denotes the discrete-time index. Furthermore, we represent signals $y[t]$, $x[t]$, and $h[t]$ in the time–frequency domain as Mel-frequency-scaled, log-magnitude spectrograms, denoted by $\mathbf{Y}[f,k]$, $\mathbf{X}[f,k]$ and $\mathbf{H}[f,k] \in \mathbb{R}^{F \times K}$, where $f$ and $k$ index frequency and time, respectively, and $F$ and $K$ denote the frequency and time dimensions of the RIR spectrogram.

\begin{figure}[t!]
    \centering
    \includegraphics[width=\columnwidth]{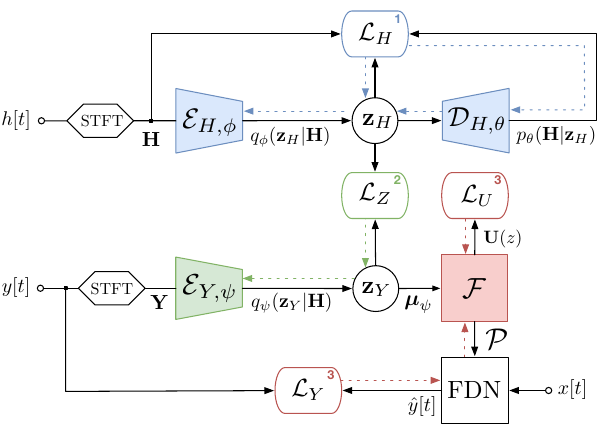}
    \caption{Overview of the proposed multi-stage approach: the loss functions are enumerated in the order in which the individual models with the corresponding colors are trained. Black arrows indicate the signal flow during the forward pass; dashed, colored arrows indicate the backpropagation of gradients.}
    \label{fig:model_overview}\vspace{-0.4cm}
\end{figure}

The proposed method involves a two-stage representation learning framework, which builds upon the approach presented in \cite{gotz2024blind}, and a final parameter estimation stage. Figure~\ref{fig:model_overview} provides a schematic overview of the complete approach, with the three colors denoting the different stages.

\subsection{Room Acoustic Prior}
 In the first stage, a variational autoencoder (VAE) \cite{kingma2013auto}, consisting of encoder $\mathcal{E}_{H,\phi}:\mathbb{R}^{F\times K}\rightarrow\mathbb{R}^{2D}$ parameterized by $\mathbf{\phi}$, and decoder $\mathcal{D}_{H,\theta}:\mathbb{R}^{D}\rightarrow\mathbb{R}^{F\times K}$ parameterized by $\mathbf{\theta}$, is trained to learn a compact representation of RIRs with $D\ll FK$. We jointly learn the posterior and likelihood distributions, $q_\phi(\mathbf{z}_H|\mathbf{H}) \sim \mathcal{N}(\bm{\mu}_\phi,\bm{\Sigma}_\phi)$ and $p_\theta(\mathbf{H}|\mathbf{z}_H)$, respectively, where $\bm{\Sigma}_\phi = \operatorname{diag}(\sigma_\phi^{2,(1)}, \dots, \sigma_\phi^{2,(D)})$. We optimize the well-known evidence lower bound objective \cite{kingma2013auto}:
\begin{equation}\label{eq:elbo}
\begin{split}
    \mathcal{L}_H(\mathbf{\phi}, \mathbf{\theta}, \mathbf{H}) = &\;\mathbb{E}_{p(\mathbf{H})}\bigg[ \lambda\mathrm{KL}\left\{q_\phi(\mathbf{z}_H|\mathbf{H})\,||\,p(\mathbf{z}) \right\} \\
    & -(1-\lambda)\mathbb{E}_{q_\phi\left(\mathbf{z}_H|\mathbf{H}\right)}\big[\log p_\theta(\mathbf{H}|\mathbf{z}_H)\big]\bigg],
\end{split}
\end{equation}
where $p(\mathbf{z})\sim\mathcal{N}(\mathbf{0}_D,\mathbf{I}_{D\times D})$ is a standard normal prior, $\mathrm{KL}\left\{p||q\right\}$ denotes the Kullback-Leibler (KL) divergence between probability distributions $p$ and $q$, and $\mathbb{E}[\cdot]$ denotes statistical expectation. The encoder $\mathcal{E}_{H,\phi}$ and decoder $\mathcal{D}_{H,\theta}$ are realized through convolutional layers, and the variational bottleneck constrains the posterior mean to $(-1,1)$ and the variance to $(0,2)$ via $\tanh(\cdot)$ activation.

In the following stage, a second encoder $\mathcal{E}_{Y,\psi}:\mathbb{R}^{F\times K}\rightarrow\mathbb{R}^{D}$ is trained to approximate the RIR posterior from reverberant speech. To this end, we minimize the KL divergence between the variational distribution $q_\psi(\mathbf{z}_Y|\mathbf{H},\!\mathbf{X})$, which is conditioned on the anechoic source signal $\mathbf{X}$, and the posterior distribution of the RIR:
\begin{equation}\label{eq:KLD}
    \argmin_{\mathbf{\psi}}\mathrm{KL}\left\{ q_\psi(\mathbf{z}_Y|\mathbf{H,X})||q_\phi(\mathbf{z}_H|\mathbf{H}) \right\}.
\end{equation}
This objective enforces that $\mathcal{E}_{Y,\psi}$ learns a latent representation of reverberant speech that is invariant to the anechoic source signal, yielding an approximation of the form $q_\psi(\mathbf{z}_Y|\mathbf{H})$. In the considered scenario, we do not sample from the speech posterior and can assume a fixed identity covariance matrix $q_\psi(\mathbf{z}_Y|\mathbf{H,X})\sim\mathcal{N}(\bm{\mu}_\psi,\mathbf{I})$. Hence, the speech encoder is only required to estimate the variational mean $\bm{\mu}_\psi$, which reduces (\ref{eq:KLD}) to:
\begin{equation}
    \mathcal{L}_Z = \frac{1}{2}\Big[\operatorname{tr}(\bm{\Sigma}_\phi^{-1})] + (\bm{\mu}_\phi - \bm{\mu}_\psi)^\top\bm{\Sigma}_\phi^{-1} (\bm{\mu}_\phi - \bm{\mu}_\psi) + \log\lvert\bm{\Sigma}_\phi\rvert - D \Big],
\end{equation}
where the operator $(\cdot)^\top$ indicates transposition. $\mathcal{E}_{Y,\psi}$ is realized as a convolutional feature extractor followed by a transformer encoder with attention-based sequence aggregation, which maps the variable-length input spectrogram $\mathbf{Y}$ to a fixed-size embedding in $\mathbb{R}^D$.  

\subsection{Artificial Reverberation}
The latent approximation $\bm{\mu}_\psi$ serves as input to the parameter estimation model, which predicts a set of FDN parameters used to synthesize reverberation of the anechoic source signal $x[t]$ (cf. Fig.~\ref{fig:model_overview}). The objective in this last stage is to align the artificially reverberated signal with the true reverberant signal.

\subsubsection{Differentiable FDN}

\begin{figure}[t!]
    \centering
    \includegraphics[trim={1.1cm 0.3cm 1.1cm 0.7cm}, clip, width=\columnwidth]{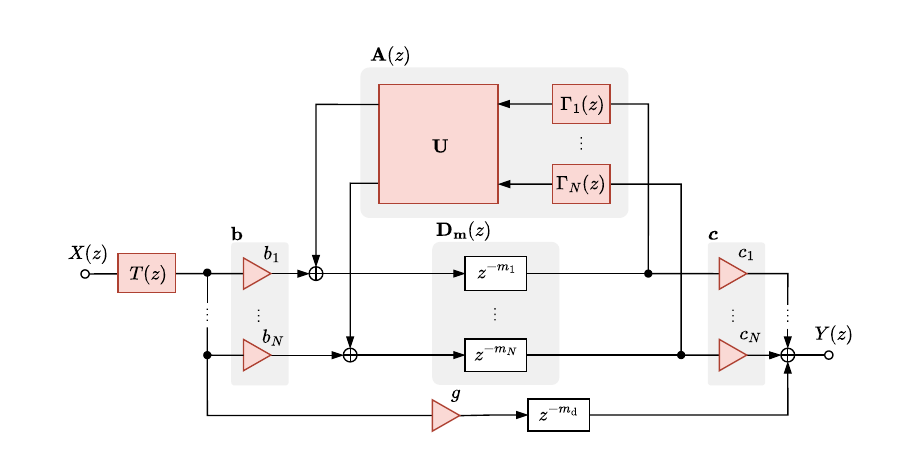}
    \caption{Proposed differentiable FDN structure. The blocks highlighted in red are estimated by the parameter estimation model. }
    \label{fig:fdn}
\end{figure}

To synthesize the reverberation, we adopt an implementation of the FDN, depicted in Fig.~\ref{fig:fdn} and defined by the following transfer function:
\begin{align}\label{eq:tr_fdn}
H(z) = T(z)\left(\mathbf{c}^\top\big[\mathbf{D_m}(z)^{-1} -\mathbf{A}(z)\big]^{-1}\mathbf{b} + gz^{-m_{\textrm{d}}}\right),
\end{align}
where $\mathbf{A}(z)$ is the filter feedback matrix, formed by combining an orthogonal $N \times N$ mixing matrix $\mathbf{U}$ with channel-wise attenuation filters $\mathbf{\Gamma}(z) = \text{diag}\{\Gamma_1(z), \dots,\Gamma_{N}(z)\}$, i.e. $\mathbf{A}(z) = \mathbf{U} \mathbf{\Gamma}(z)$. Here, $N$ denotes the number of delay lines. Each filter is a graphic equalizer (GEQ) implemented as a cascade of $J$ second-order sections, comprising shelving filters, peaking filters, and a scalar gain. The vector of delay lengths is $\mathbf{m} = [m_1, \dots, m_N]$, and the corresponding delay matrix $\mathbf{D_m}(z)$ is a diagonal matrix with entries $[z^{-m_1}, \dots, z^{-m_N}]$. Vectors $\mathbf{b}$ and $\mathbf{c}$ are $N \times 1$ column vectors representing the input and output gains, respectively. The direct path is modeled by a gain $g$ and a delay $m_{\textrm{d}}$. $T(z)$ denotes the tone-correction GEQ.
All parameters, apart from the delay lines, are learnable. 

In traditional FDN design, the attenuation filters $\mathbf{\Gamma}(z)$ are typically scaled in proportion to the delay lengths \cite{jot1992analysis} to model a single frequency-dependent decay rate. In the proposed design, however, we lift this constraint to avoid numerical instabilities during training. The tone-corrector filter $T(z)$ is necessary to model the frequency-dependent initial energy of the reference RIR. The proposed FDN structure is realized using the \textit{FLAMO}\footnote{\url{https://github.com/gdalsanto/flamo}} Python library, which provides a comprehensive toolbox for differentiable signal processing in the frequency domain based on the frequency sampling method \cite{dal2025flamo}.

\subsubsection{FDN Parameter Estimation}
We estimate the FDN parameters using a set of regression models $\mathcal{F}$, each consisting of a shallow multi-layer perceptron (MLP) that maps the latent representation $\bm{\mu}_\psi$ to a specific FDN parameter. $\mathcal{F}:\mathbb{R}^D\rightarrow \mathcal{P}$ yields the parameter set:
\begin{equation}
\mathcal{P} = \left\{
\begin{array}{ll}
p_T \in \mathbb{R}^{J \times 1}, & \mathbf{b} \in \mathbb{R}^{N \times 1}, \\
p_\mathbf{U} \in \mathbb{R}^{N \times N}, & \mathbf{c} \in \mathbb{R}^{N \times 1}, \\
p_{\bm{\Gamma}} \in \mathbb{R}^{J \times N}, & g \in \mathbb{R}
\end{array}
\right\},
\end{equation}
which is used to reverberate the dry signal $x[t]$, obtaining $\hat{y}[t]$. In this work, we use $N=8$ and $J=11$.

To guarantee FDN stability, the intermediate parameters $p_T$, $p_\mathbf{U}$, and $p_{\bm{\Gamma}}$ are transformed into the corresponding FDN components $T(z)$, $\mathbf{U}$, and $\mathbf{\Gamma}(z)$ through dedicated activation functions. $\mathbf{U}$ is obtained using the orthogonality mapping $\mathbf{U} = \exp\{ \textrm{Tr}(p_{\mathbf{U}}) - \textrm{Tr}(p_{\mathbf{U}})^\top\}$, where $ {\textrm{Tr}}(\cdot)$ extracts the upper triangular matrix and the operator $\exp\{ \cdot  \}$ denotes the matrix exponential \cite{lezcano2019cheap}. Parameters $p_T$ and $p_{\bm{\Gamma}}$ control the command gains of the one-octave band GEQs and are limited to $[-12, 12]\,$dB and $[-\inf, 0]\,$dB, respectively, using a scaled $\tanh(\cdot)$ activation and a $\text{sigmoid}(\cdot)$ activation. The chosen GEQ design at the 48-kHz sample rate has nine peaking stages with center frequencies ranging from 62.5~Hz to 16~kHz, two shelving stages with cutoff frequencies of 44~Hz and 22.6~kHz, and a dc gain factor. 
The lengths of the delay lines are fixed to $\mathbf{m} = \left[809, 877, 937, 1049, 1151, 1249, 1373, 1499\right]$. The values in $\mathbf{m}$ are coprime numbers distributed logarithmically, aiming to maximize the echo density \cite{schlecht2016feedback} and avoid degenerative patterns. Since the silence before the onset is removed during data pre-processing, the direct path is modeled with a short two-sample delay $m_\textrm{d}$, and the direct-to-reverberant ratio is controlled by $g$.
 
We train $\mathcal{F}$ with a multi-resolution STFT loss function \cite{yamamoto2020parallel,steinmetz2020auraloss}, defined as the mean squared distance between the Mel-frequency scaled, logarithmic-magnitude spectra of $y[t]$ and $\hat{y}[t]$:
\begin{align}
\mathcal{L}_{Y}(y,\hat{y})
\;=&\;
\frac{1}{|\mathcal{R}|}\sum_{r\in\mathcal{R}}
\frac{1}{F_r T_r}\;
\big\|\, \mathbf{Y}^{(r)} - \widehat{\mathbf{Y}}^{(r)}\,\big\|_F^{2}\\
\text{with}\;\; \mathbf{Y}^{(r)}(y) \;\triangleq&\; 10\log_{10}\!\left(\mathbf{M}_r\,\big|\mathrm{STFT}_{r}\{y\}\big|^2 + \varepsilon\right),
\end{align}
where $\mathbf{M}$ denotes a mapping from linear to Mel-frequency scale, $\mathcal{R}$ is a set of spectrogram configurations, each with parameters $\left(N_{\mathrm{fft}}^{(r)}, N_{\mathrm{hop}}^{(r)}, N_{\mathrm{mel}}^{(r)}\right)$, $|\cdot|$ indicates cardinality, $\lVert\cdot\rVert_F$ denotes the Frobenius norm, and $\varepsilon>0$ is a constant preventing $\log(0)$. In addition to the signal matching objective, we enforce a dense feedback matrix $\mathbf{U}$ with a sparsity penalty, which encourages a fast build-up of temporal reflection density at the beginning of the FDN response and yields a perceptually smooth reverberation tail \cite{dal2024rir2fdn}:
\begin{equation}
    \mathcal{L}_U=\frac{N\sqrt{N}-\sum_{i,j}\lvert U_{i,j}\rvert}{N\left(\sqrt{N}-1\right)},
\end{equation}
where $U_{i,j}$ is the entry of matrix $\mathbf{U}$ in the $i$-th row and $j$-th column. We train the parameter estimator $\mathcal{F}$ with a weighted sum of both loss terms $\mathcal{L}_3=\mathcal{L}_Y+\lambda \mathcal{L}_U$ (cf. Fig.~\ref{fig:model_overview}).

\section{Evaluation Setup} \label{sec:eval} 
\vspace{-0.5em}
\subsection{Dataset}
We generate a large corpus of reverberant speech by combining anechoic speech from the EARS dataset \cite{richter2024ears} with RIRs from a variety of publicly available datasets, including ACE \cite{eaton2016estimation}, ASH-IR \cite{ash_dataset}, Multi-Room Transition \cite{goetz_blind_slopes}, Arni \cite{karolina2022dataset}, EchoThief \cite{echothief}, IKS \cite{jeub2009binaural}, MIT \cite{traer2016statistics}, OpenAir \cite{murphy2010openair}, Multi-Purpose Room Impulse Response \cite{friede2024multipurpose}, TAU spatial RIR \cite{politis2022tau}, and TH Köln \cite{lubeck2021high}. We partition all RIRs and speech signals into three disjoint sets for training, validation, and testing, and generate approximately $18$ h of 4-s reverberant speech segments, sampled at $48\,\text{kHz}$. RIRs are energy-normalized with onset times removed, and spectrograms are standardized to zero mean and unit variance.

\vspace{-0.5em}
\subsection{Model and Training Details}
The VAE used to learn the RIR posterior distribution comprises \num{393857} parameters, the speech encoder for the latent approximation comprises \num{1475168} parameters, and the FDN parameter estimator comprises \num{573465} parameters. Hence, during inference, we run a model with a total of \num{2048444} learnable parameters. We train all models using the Adam optimizer with decoupled weight decay \cite{loshchilov2017decoupled} and a learning rate schedule. We then select the model with the lowest validation loss after a fixed patience period of $16$ epochs.

\vspace{-0.5em}
\subsection{Evaluation Metrics}

\begin{figure*}[ht]
    \centering
    \includegraphics[width=\textwidth]{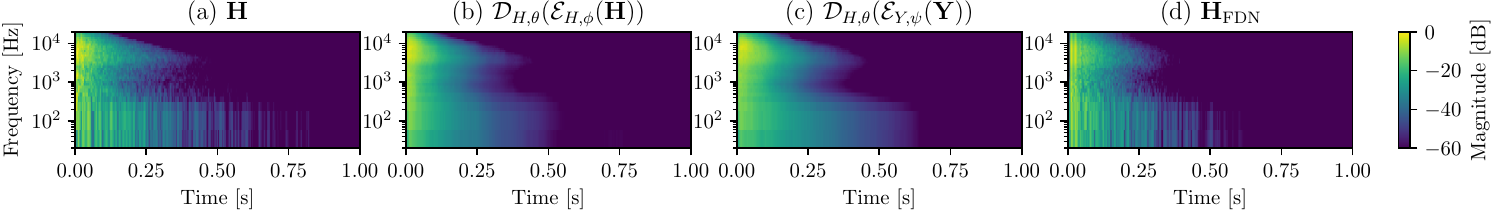}
    \caption{Qualitative results of the proposed method: (a) RIR spectrogram, (b) VAE reconstruction, (c) decoded approximation from reverberant speech, and (d) synthesized FDN response spectrogram.}
    \label{fig:spectrograms}\vspace{-3mm}
\end{figure*}

We evaluate the proposed method by comparing acoustic parameters between true and artificial RIRs. Specifically, \tthirty and \cfifty are assessed across seven octave bands with center frequencies $\{125, 250, 500, 1\mathrm{k}, 2\mathrm{k}, 4\mathrm{k}, 8\mathrm{k}\}\,$Hz, reporting the mean absolute percentage error for \tthirty, the mean absolute error for \cfifty, and the Pearson correlation coefficient (PCC) for both parameters. In addition to the parametric evaluation, we assess the perceptual plausibility of the artificially reverberant speech reflected by the FAD \cite{kilgour2019frechet}.

\subsection{Baseline}
At the time of writing, the approach proposed by Lee et al.~\cite{lee2022differentiable} represents the leading method in blind estimation of FDN parameters from reverberant speech. We therefore adopt it as our baseline method. In \cite{lee2022differentiable}, the encoder and projection layer are trained by minimizing a linear multi-resolution spectral loss. The encoder consists of five 2D convolutional layers, two gated recurrent units, and two linear layers. The projection layers consist of two linear layers followed by the parameter-specific activation function. 

The FDN structure for which ARP-net estimates parameters has size $N=6$ and includes learnable $\mathbf{b}$ and $\mathbf{c}$, a fixed Householder feedback matrix $\mathbf{U}$, learnable attenuation filters with a common response $\Gamma(z)$, a common tone-correction filter $T(z)$, and a cascade of four Schroeder allpass (SAP) filter sections in each feedback path with learnable gains. The SAPs were deemed necessary to achieve faster echo density build-up without adding more delay lines \cite{lee2022differentiable}. As in our proposed method, all delay lines are fixed at initialization. The filters are implemented as eight-stage parametric equalizers.

To synthesize the energy decay of the reference RIR, Lee et al.~\cite{lee2022differentiable} employ a common absorption filter $\Gamma(z)$, implemented as an eight-stage parametric equalizer using state-variable filter (SVF) parameters. This filter consists of one low-shelving, six peaking, and one high-shelving filter. ARP-net is trained to estimate the resonance, cutoff frequencies, and gains of each band. Allowing the cutoff frequency to vary for each RIR improves the network’s generalization ability. The tone-correction filter $T(z)$ is implemented similarly as a series of eight SVF filters, each with a learnable cutoff frequency, resonance, and mixing coefficients.

\begin{table}[b!]
 \caption{Comparison of artificially reverberated speech generated by the proposed approach and the baseline in terms of FAD.}
    \centering
    \begin{tabular}{c c c}
    \toprule
         & Proposed method & ARP-net \cite{lee2022differentiable} \\
         \hline
         FAD \cite{kilgour2019frechet} ($\downarrow$) & $0.109$ & $0.523$ \\
         \bottomrule
    \end{tabular}
   
    \label{tab:fad}
\end{table}

\section{Results}
\label{sec:results}
\begin{figure}[t!]
    \centering
    \includegraphics[width=\columnwidth]{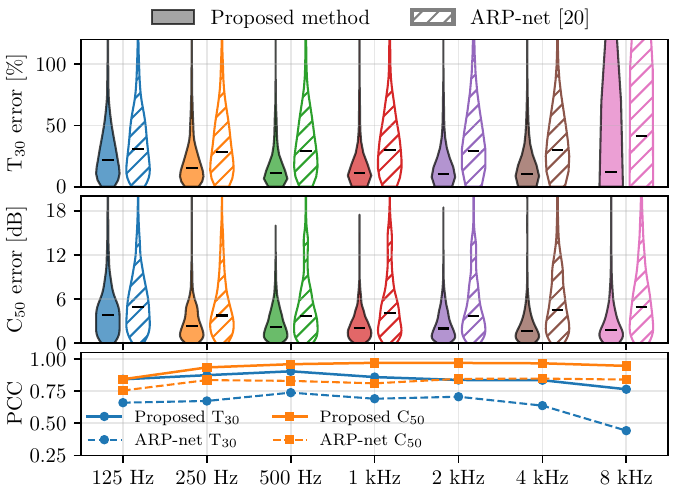}
    \caption{Comparison of the proposed approach and the baseline in terms of acoustic parameters of the synthesized RIRs. For each octave band, the top two plots show empirical distributions of parameter errors (lower is better), and the bottom plot shows the PCC between ground-truth and synthesized parameters (higher is better).}\vspace{-5mm}
    \label{fig:results}
\end{figure}

The empirical error distributions shown in the top two plots of Fig.~\ref{fig:results} indicate that the proposed method accurately matches both acoustic parameters, outperforming the baseline approach. For the proposed method, optimal performance is observed in the frequency range between $500$~Hz and $2$~kHz, where the concentration of speech energy yields the highest signal-to-noise ratio and, consequently, supports accurate estimation of the acoustic prior. The PCCs between the parameters of the ground truth and the synthesized RIRs, shown in the bottom plot of Fig.~\ref{fig:results}, highlight the advantage of the proposed method, which is most pronounced for \tthirty. A general trend observed for both methods is that estimating acoustic parameters is most challenging in the lowest and highest octave bands, where speech does not sufficiently excite the acoustic response.

The perceptual quality of the synthesized reverberant speech is evaluated using the FAD, as reported in Table \ref{tab:fad}. While a formal listening experiment would be required to directly assess human preference, the results demonstrate that the proposed approach is capable of generating highly plausible artificially reverberated speech signals\footnote{Sound examples: \url{https://www.audiolabs-erlangen.de/resources/2026-ICASSP-RMS}}. Finally, we highlight the considerable modularity and control afforded by the multi-stage design of the proposed method, as illustrated in Fig.~\ref{fig:spectrograms} with a single RIR example. This design enables the assessment of both the level of detail captured in the acoustic prior and the extent to which this prior can be estimated from a given segment of reverberant speech, on which the quality of the synthesized RIR ultimately depends.

\vspace{-1mm}
\section{Conclusion}
\label{sec:conclusion}
We have presented a method for the blind estimation of FDN parameters from reverberant speech. The proposed approach extracts a room-acoustic prior and employs a differentiable FDN structure capable of modeling both frequency-dependent decay times and the direct-to-reverberant ratio. Evaluation of the synthesized RIRs, in terms of acoustic parameters, together with the perceptual plausibility of the generated reverberant speech, demonstrated that the proposed method outperformed the baseline.
\clearpage

\bibliographystyle{IEEEbib_twoauth}
\bibliography{bibliography}

\begin{thebibliography}{10}

\bibitem{blauert1997spatial}
J.~Blauert,
\newblock {\em {Spatial Hearing: The Psychophysics of Human Sound
  Localization}},
\newblock MIT Press, Cambridge, MA, USA, 1997.

\bibitem{bregman1994auditory}
A.~S. Bregman,
\newblock {\em {Auditory Scene Analysis: The Perceptual Organization of
  Sound}},
\newblock MIT Press, Cambridge, MA, USA, 1994.

\bibitem{agrawal2019defining}
S.~Agrawal et~al.,
\newblock ``{Defining immersion: Literature review and implications for
  research on immersive audiovisual experiences},''
\newblock {\em J. Audio Eng. Soc.}, vol. 67, no. 11, pp. 886--897, 2019.

\bibitem{neidhardt2022perceptual}
A.~Neidhardt et~al.,
\newblock ``{Perceptual matching of room acoustics for auditory augmented
  reality in small rooms-literature review and theoretical framework},''
\newblock {\em Trends Hear.}, vol. 26, Apr. 2022.

\bibitem{potter2022relative}
T.~Potter et~al.,
\newblock ``{On the relative importance of visual and spatial audio rendering
  on VR immersion},''
\newblock {\em Front. Signal Process.}, vol. 2, May 2022.

\bibitem{immohr2023proof}
F.~Immohr et~al.,
\newblock ``{Proof-of-concept study to evaluate the impact of spatial audio on
  social presence and user behavior in multi-modal VR communication},''
\newblock in {\em Proc. ACM IMX}, Jun. 2023, pp. 209--215.

\bibitem{meyer2024testing}
N.~Meyer-Kahlen et~al.,
\newblock ``Testing auditory illusions in augmented reality: Plausibility,
  transfer-plausibility, and authenticity,''
\newblock {\em J. Audio Eng. Soc.}, vol. 72, no. 11, pp. 797--812, 2024.

\bibitem{kuttruff2016room}
H.~Kuttruff,
\newblock {\em {Room Acoustics}},
\newblock CRC Press, 2016.

\bibitem{eaton2016estimation}
J.~Eaton et~al.,
\newblock ``{Estimation of room acoustic parameters: The ACE challenge},''
\newblock {\em IEEE/ACM Trans. Audio, Speech, Lang. Process.}, vol. 24, no. 10,
  pp. 1681--1693, Oct. 2016.

\bibitem{lin2006bayesian}
Y.~Lin et~al.,
\newblock ``{Bayesian regularization and nonnegative deconvolution for room
  impulse response estimation},''
\newblock {\em IEEE Trans. Signal Process.}, vol. 54, no. 3, pp. 839--847, Mar.
  2006.

\bibitem{crammer2006room}
K.~Crammer et~al.,
\newblock ``{Room impulse response estimation using sparse online prediction
  and absolute loss},''
\newblock in {\em Proc. ICASSP}, May 2006, vol.~3, pp. 301--304.

\bibitem{steinmetz2021filtered}
C.~J. Steinmetz et~al.,
\newblock ``{Filtered noise shaping for time domain room impulse response
  estimation from reverberant speech},''
\newblock in {\em Proc. WASPAA}, Oct. 2021, pp. 221--225.

\bibitem{lee2023yet}
S.~Lee et~al.,
\newblock ``{Yet another generative model for room impulse response
  estimation},''
\newblock in {\em Proc. WASPAA}, Oct. 2023.

\bibitem{ratnarajah2023towards}
A.~Ratnarajah et~al.,
\newblock ``{Towards improved room impulse response estimation for speech
  recognition},''
\newblock in {\em Proc. ICASSP}, Jun. 2023.

\bibitem{moliner2024buddy}
E.~Moliner et~al.,
\newblock ``{BUDDy: Single-channel blind unsupervised dereverberation with
  diffusion models},''
\newblock in {\em Proc. IWAENC}, Sep. 2024, pp. 120--124.

\bibitem{schlecht2016feedback}
S.~J. Schlecht et~al.,
\newblock ``{Feedback delay networks: Echo density and mixing time},''
\newblock {\em IEEE/ACM Trans. Audio, Speech, Lang. Process.}, vol. 25, no. 2,
  pp. 374--383, Feb. 2017.

\bibitem{gerzon1971synthetic}
M.~A. Gerzon,
\newblock ``{Synthetic stereo reverberation: Part one},''
\newblock {\em Studio Sound}, vol. 13, no. 12, pp. 632--635, Dec. 1971.

\bibitem{jot1991digital}
J.-M. Jot et~al.,
\newblock ``{Digital delay networks for designing artificial reverberators},''
\newblock in {\em Proc. AES Conv.}, Feb. 1991.

\bibitem{valimaki2012fifty}
V.~V{\"a}lim{\"a}ki et~al.,
\newblock ``{Fifty years of artificial reverberation},''
\newblock {\em IEEE Trans. Audio, Speech, Lang. Process.}, vol. 20, no. 5, pp.
  1421--1448, Jul. 2012.

\bibitem{lee2022differentiable}
S.~Lee et~al.,
\newblock ``{Differentiable artificial reverberation},''
\newblock {\em IEEE/ACM Trans. Audio, Speech, Lang. Process.}, vol. 30, pp.
  2541--2556, Aug. 2022.

\bibitem{dal2024rir2fdn}
G.~Dal~Santo et~al.,
\newblock ``{RIR2FDN: An improved room impulse response analysis and
  synthesis},''
\newblock in {\em Proc. DAFx}, Sep. 2024, pp. 230--237.

\bibitem{kilgour2019frechet}
K.~Kilgour et~al.,
\newblock ``{Fr\'echet Audio Distance: A} metric for evaluating music
  enhancement algorithms,''
\newblock in {\em Proc. Interspeech}, Sep. 2019, pp. 2350--2354.

\bibitem{gotz2024blind}
P.~G{\"o}tz et~al.,
\newblock ``{Blind acoustic parameter estimation through task-agnostic
  embeddings using latent approximations},''
\newblock in {\em Proc. IWAENC}, Sep. 2024, pp. 289--293.

\bibitem{kingma2013auto}
D.~P. Kingma et~al.,
\newblock ``{Auto-encoding variational Bayes},''
\newblock {\em arXiv preprint arXiv:1312.6114}, Dec. 2013.

\bibitem{jot1992analysis}
J.-M. Jot,
\newblock ``{An analysis/synthesis approach to real-time artificial
  reverberation},''
\newblock in {\em Proc. ICASSP}, 1992, pp. 221--224.

\bibitem{dal2025flamo}
G.~Dal~Santo et~al.,
\newblock ``{FLAMO: An} open-source library for frequency-domain differentiable
  audio processing,''
\newblock in {\em Proc. ICASSP}, Apr. 2025.

\bibitem{lezcano2019cheap}
M.~Lezcano-Casado et~al.,
\newblock ``{Cheap orthogonal constraints in neural networks: A simple
  parametrization of the orthogonal and unitary group},''
\newblock in {\em Proc. ICML}, May 2019, pp. 3794--3803.

\bibitem{yamamoto2020parallel}
R.~Yamamoto et~al.,
\newblock ``{Parallel WaveGAN: A fast waveform generation model based on
  generative adversarial networks with multi-resolution spectrogram},''
\newblock in {\em Proc. ICASSP}, May 2020, pp. 6199--6203.

\bibitem{steinmetz2020auraloss}
C.~J. Steinmetz et~al.,
\newblock ``{Auraloss: Audio focused loss functions in PyTorch},''
\newblock in {\em Proc. DMRN+15}, Dec. 2020.

\bibitem{richter2024ears}
J.~Richter et~al.,
\newblock ``{EARS}: An anechoic fullband speech dataset benchmarked for speech
  enhancement and dereverberation,''
\newblock in {\em Proc. Interspeech}, Sep. 2024.

\bibitem{ash_dataset}
S.~Pearce,
\newblock ``{ASH-IR dataset},''
  \url{https://github.com/ShanonPearce/ASH-IR-Dataset},
\newblock Accessed: 2024-09-26.

\bibitem{goetz_blind_slopes}
P.~G{\"o}tz et~al.,
\newblock ``A multi-room transition dataset for blind estimation of energy
  decay,''
\newblock in {\em Proc. IWAENC}, Sep. 2024, pp. 125--129.

\bibitem{karolina2022dataset}
K.~Prawda et~al.,
\newblock ``{Dataset of impulse responses from variable acoustics room Arni at
  Aalto Acoustic Labs},'' Zenodo, \url{10.5281/zenodo.6582103}, 2022.

\bibitem{echothief}
C.~Warren,
\newblock ``{EchoThief impulse response library},''
  \url{http://www.echothief.com/},
\newblock Accessed: 2024-05-14.

\bibitem{jeub2009binaural}
M.~Jeub et~al.,
\newblock ``{A binaural room impulse response database for the evaluation of
  dereverberation algorithms},''
\newblock in {\em Proc. DSP}, Jul. 2009.

\bibitem{traer2016statistics}
J.~Traer et~al.,
\newblock ``{Statistics of natural reverberation enable perceptual separation
  of sound and space},''
\newblock {\em Proc. Nat. Acad. Sci.}, vol. 113, no. 48, pp. E7856--E7865, Nov.
  2016.

\bibitem{murphy2010openair}
D.~T. Murphy et~al.,
\newblock ``{Openair: An interactive auralization web resource and database},''
\newblock in {\em Proc. AES Conv.}, May 2010.

\bibitem{friede2024multipurpose}
L.~Friede et~al.,
\newblock ``Multi-purpose room impulse response dataset measured on a {3D}
  spatial grid,''
\newblock in {\em Proc. AES Conv.}, Jun. 2024.

\bibitem{politis2022tau}
A.~Politis et~al.,
\newblock ``{TAU spatial room impulse response database (TAU-SRIR DB)},''
  Zenodo, \url{https://zenodo.org/record/6408611}, Apr. 2022.

\bibitem{lubeck2021high}
T.~L{\"u}beck et~al.,
\newblock ``A high-resolution spatial room impulse response database,''
\newblock in {\em Proc. 47th DAGA}, Aug. 2021.

\bibitem{loshchilov2017decoupled}
I.~Loshchilov et~al.,
\newblock ``Decoupled weight decay regularization,''
\newblock in {\em Proc. ICLR}, May 2019.

\end{thebibliography}

\end{document}